\documentstyle[12pt]{article}

\renewcommand{\baselinestretch}{1.59}

\newtheorem{theorem}{Theorem}            

\begin{document}

\renewcommand{\baselinestretch}{1.1}

\title{\Large \bf On a Class of Riemann-Cartan Space-times 
                  of G\"odel Type \\} 
\author{
J.B. Fonseca-Neto\thanks{
{\sc internet: joel@dfjp.ufpb.br}  } \ \ 
and  \  
M.J. Rebou\c{c}as\thanks{
{\sc internet: reboucas@cat.cbpf.br}  }  \\ 
\\  
$~^{\ast}~$Departamento de F\'{\i}sica,
             Universidade Federal da Para\'{\i}ba  \\
             Caixa Postal 5008,\  58059-900 Jo\~ao Pessoa -- PB, Brazil \\   
\\
$^{\dagger}$~Centro Brasileiro de Pesquisas F\'\i sicas\\
             Departamento de Relatividade e Part\'\i culas \\
             Rua Dr.\ Xavier Sigaud 150 \\
             22290-180 Rio de Janeiro -- RJ, Brazil \\ \\
       }
        
\date{\today}

\maketitle

\begin{abstract}
A class of Riemann-Cartan G\"odel-type space-times is examined 
by using the equivalence problem techniques, as formulated by 
Fonseca-Neto {\em et al.\/}
and embodied in a suite of computer algebra programs 
called {\sc tclassi}. 
A coordinate-invariant description of the gravitational field
for this class of space-times is presented.
It is also shown that these space-times can admit a group $G_{r}$ of 
affine-isometric motions of dimensions $r=2, 4, 5$.
The necessary and sufficient conditions for space-time (ST) 
homogeneity of this class of space-times are derived, 
extending previous works on G\"odel-type space-times.
The equivalence of space-times in the ST homogeneous subclass 
is studied, recovering recent results 
under different premises. The results of the limiting 
Riemannian case are also recovered.
\end{abstract}

\newpage

{\raggedright
 \section{Introduction} }      \label{intro}
\setcounter{equation}{0}

The G\"{o}del solution of Einstein's field equations~\cite{godel} 
is a particular case of the G\"{o}del-type line element
\begin{equation}
ds^{2} = [ dt + H(x)\, dy]^{2} - D^{2}(x) \, dy^{2} - dx^{2} - dz^{2},
                                                        \label{ds2}
\end{equation}
in which 
\begin{equation}
  H(x) = e^{m x}, \;\;\;\;  D(x) = e^{m x}/ \sqrt{2},    \label{gddgod}
\end{equation}
where $m$ is a real constant, which is related to the cosmological 
constant $\Lambda$ and the rotation $\omega$ by  
$m^{2} = 2\, \omega^{2} = - 2\, \Lambda$. 
The G\"{o}del model is homogeneous in space-time (ST homogeneous), 
since it admits a five-dimensional isometry group $G_{5}$, with
an isotropy subgroup of dimension one ($H_{1}$).

Despite its striking properties, the cosmological solution
presented by G\"odel has a well recognized historical (and
even philosophical~\cite{Pfarr81}) importance and has to a
large extent motivated the investigation on rotating cosmological
space-times. Particularly, the search for rotating G\"odel-type
space-times has received a great deal of attention in recent 
years, and the literature on these geometries is fairly large
today (see, for example, Krasi\'nski~\cite{Krasinski97},
Singh and Agrawal~\cite{SinghAgrawal94}, and 
references therein).

In general relativity (GR), the space-time $M$ is a 
four-dimensional {\em Riemannian\/} manifold $M$ endowed with a 
locally Lorentzian metric $g_{ab}$. In GR there is a unique
metric-compatible symmetric connection 
$\{_{b\ c}^{\ a}\}$ (Christoffel's symbols). 
However, it is well known that the metric tensor and the 
connection can be introduced as independent structures on a 
given space-time manifold $M$. 
In the framework of torsion theories of gravitation (TTG),
we have {\em Riemann-Cartan\/} (RC) manifolds, i.e., space-time 
manifolds endowed with locally Lorentzian metrics $g_{ab}$ 
and metric-compatible nonsymmetric connections 
${\Gamma}^a_{\ bc}$. Thus, in TTG the connection has a 
metric-independent part given by the torsion, and for a 
characterization of the local gravitational field one has 
to deal with both metric and connection. 

In GR and TTG the arbitrariness in the choice of coordinates 
gives rise to the problem of deciding whether or not two 
apparently different space-time solutions of the field equations 
are locally the same (the equivalence problem). In GR this 
problem can be couched in terms of local isometry, whereas in TTG 
besides local isometry ($g_{ab} \rightarrow \tilde{g}_{ab}$) 
it means local affine collineation~%
($\Gamma^{a}_{\ bc} \rightarrow \tilde{\Gamma}^{a}_{\ bc}$) 
of two RC manifolds.

The local equivalence for the Riemannian space-times of general 
relativity  has been discussed by several authors and is of 
interest in many contexts~\cite{cartan}~--~\cite{MacCSkea}.
The conditions for the local equivalence of Riemann-Cartan 
space-times in the torsion theories of gravitation, however,
have been found only recently~\cite{frt}. 
Subsequently,  an algorithm for checking the equivalence 
in TTG  and  a working version of a computer algebra 
package (called {\sc tclassi}) which implements this algorithm 
have been presented~\cite{frm}~--~\cite{afmr}.

The problem of space-time homogeneity (ST homogeneity) of 
four-dimensional {\em Riemannian\/} manifolds endowed with a 
G\"{o}del-type metric (\ref{ds2}) was considered for the first 
time by Raychaudhuri and Thakurta~\cite{raytha} in 1980. They 
have shown that the conditions
\begin{eqnarray}
\frac{H'}{D} &=&\mbox{const} \equiv 2\,\omega \label{metcond1} \,, \\
\frac{D''}{D}&=&\mbox{const} \equiv m^2  \label{metcond2}
\end{eqnarray}
are necessary for a Riemannian G\"{o}del-type space-time manifold
to be ST-homogeneous. Here and in what follows we use the prime
to denote derivative with respect to $x$.
In 1983, it was proved~\cite{rebtio} that the conditions 
(\ref{metcond1}) and (\ref{metcond2}) are also sufficient for 
ST homogeneity of Riemannian G\"odel-type space-time manifolds. 

However, in both articles~\cite{raytha,rebtio} in the study of
ST homogeneity it was explicitly or implicitly assumed that
G\"odel-type space-time manifolds can admit only time-independent 
Killing vector fields~\cite{tra}. 
The conditions (\ref{metcond1}) and (\ref{metcond2}) were
finally proved to be the necessary and sufficient conditions for a 
Riemannian G\"{o}del-type  space-time manifold to be ST homogeneous 
in a more general setting in~\cite{rebaman}, where the 
powerful equivalence problem techniques for Riemannian 
space-times, as formulated by Karlhede~\cite{karl} and implemented 
in the computer algebra package {\sc classi}~\cite{Aman} were used.

In a recent work {\AA}man {\em et al.\/}%
~\cite{AmanFonsecaMacCallumReboucas97} 
have used the equivalence problem techniques for TTG 
to study the Riemann-Cartan manifolds endowed with a G\"odel-type 
metric (\ref{ds2}) and a torsion  $T^{t}_{\ xy}=D(x)\,S(x)$ in the 
same coordinate system relative to which the metric (\ref{ds2})
is given.  Since in the context of Einstein-Cartan theory a torsion 
with only this nonvanishing component corresponds to a Weyssenhoff 
fluid whose vector associated to the spin density is aligned along the 
direction of the rotation vector (the z axis)~\cite{OTT1,OTT2}, 
through out this article we shall refer to this torsion as 
polarized (aligned) along the rotation vector. Clearly this torsion 
also shares the same translational symmetries of the 
metric~(\ref{ds2}).   

In this work, in the light of the equivalence techniques, as
formulated by Fonseca-Neto {\em et al.\/}~\cite{frt} and
embodied in the suite of computer algebra programs 
{\sc tclassi}~\cite{frm}~--~\cite{afmr}, we extend the 
above-mentioned investigations by examining a class of 
Riemann-Cartan G\"odel-type space-times in which the torsion, 
although polarized along the direction of the rotation, does not 
share the same translational symmetries of the metric (\ref{ds2}). 
A coordinate-invariant description of the gravitational field
for this class of RC space-times is presented.
We show that these RC space-times admit a group $G_{r}$ of 
affine-isometric motions with dimensions $r=2$ [when $H(x), D(x)$ 
and $S(z)$ are arbitrary smooth real functions], $r=4$ 
(when the conditions for ST homogeneity of the corresponding 
Riemannian case hold), and $r=5$ [when besides 
the conditions (\ref{metcond1})~--~(\ref{metcond2}) one has a 
constant torsion].
We also show that the RC G\"odel-type space-times that allow
a $G_4$ of symmetry are not ST homogeneous. Actually the orbit
of an arbitrary point $P$ under the action of $G_4$ in this class 
of manifolds is a three-dimensional hypersurface, and $G_4$ 
admits a subgroup of isotropy $H_1$.
It emerges from our results that the necessary and sufficient 
conditions for space-time ST homogeneity found in%
~\cite{AmanFonsecaMacCallumReboucas97} 
also hold for the class of Riemann-Cartan G\"odel-type 
space-times in which the torsion does not share the same
translational symmetries of the metric.

Our major aim in the next section is to present a summary of some 
important prerequisites for Section 3, to set our framework, 
define the notation, and make our text to a certain extent
clear and self-contained.

In Section 3 we present our main results and conclusions in 
connection with earlier results on G\"odel-type space-time manifolds%
~\cite{raytha,rebtio,rebaman,AmanFonsecaMacCallumReboucas97}.

\vspace{5mm}

{\raggedright
\section{Theoretical and Practical Preliminaries } } 
\label{pre}
\setcounter{equation}{0} 

A solution to the equivalence problem for Riemann-Cartan 
manifolds can be summarized as follows~\cite{frt,frm}. Two
$n$-dimensional Riemann-Cartan manifolds $M$ and $\widetilde{M}$ are 
locally equivalent if there exist coordinate and Lorentz
transformations such that the following equations relating the 
Lorentz frame components of the curvature and torsion tensors 
and their covariant derivatives:
\begin{eqnarray}  
\label{eqvcond} 
T^{A}_{\ BC}    & = & \widetilde{T}^{A}_{\ BC}\,, \nonumber \\
R^{A}_{\ BCD} & = &  \widetilde{R}^{A}_{\ BCD}\,, \nonumber \\
T^{A}_{\ BC;M_{1}}  & = &  \widetilde{T}^{A}_{\ BC;M_{1}}\,, \nonumber \\ 
R^{A}_{\ BCD;M_{1}} & = & \widetilde{R}^{A}_{\ BCD;M_{1}}\,, \nonumber  \\
T^{A}_{\ BC;M_{1}M_{2}}  & = &  \widetilde{T}^{A}_{\ BC;M_{1}M_{2}}\,,  \\ 
                  & \vdots &   \nonumber \\
R^{A}_{\ BCD;M_{1}\ldots M_{p+1}} & = & \widetilde{R}^{A}_{\ BCD;M_{1}
                                             \ldots M_{p+1}}\,,\nonumber \\ 
T^{A}_{\ BC ;M_{1} \ldots M_{p+2}} & = & \widetilde{T}^{A}_{\ BC;M_{1} 
                                   \ldots M_{p+2}} \nonumber 
\end{eqnarray} 
are compatible as {\em algebraic} equations in  
$\left( x^{a}, \xi^{A} \right)$ and 
$\left(\tilde{x}^{a}, \tilde{\xi}^{A} \right)$. Here 
and in what follows we use a semicolon to denote covariant derivatives.
Note that $x^{a}$ and $\tilde{x}^{a}$ are coordinates on the manifolds 
$M$ and $\widetilde{M}$, respectively, while $ \xi^{A}$ and $\tilde{\xi}^{A}$
parametrize the corresponding groups of allowed frame transformations. 
Reciprocally, equations (\ref{eqvcond}) imply local equivalence 
between the space-time manifolds.

In practice, the coordinates and Lorentz transformations
parameters are treated differently. 
Actually a fixed frame is chosen to perform the calculations 
so that only coordinates appear in the components of
the curvature and the torsion tensors; there is no explicit 
dependence on the Lorentz parameters.	

It is worth noting that in calculating the covariant derivatives 
of the curvature and torsion tensors one can stop as soon as one 
reaches a step of the differentiation process at which the $p^{th}$ 
derivatives (say) are algebraically
expressible in terms of the previous ones, since further differentiation
will not yield any new piece of information. Actually, if the $p^{th}$
derivative is expressible in terms of its predecessors, for any 
$q > p$ the $q^{th}$ derivatives can all be expressed in terms of the
$0^{th}$, $1^{st}$, $\cdots$, $(p-1)^{th}$ derivatives. As in the worst 
case we have only one functionally independent function in each step of
the differentiation process, it follows that for  four-dimensional
Riemann-Cartan manifolds $p+1 \leq 10$.

An important practical point to be considered, once one wishes to test
the local equivalence of two Riemann-Cartan manifolds, is that before
attempting to solve eqs.\ (\ref{eqvcond}) one can extract and compare
partial pieces of information as, for example, the subgroup $H_q$ of
the symmetry group $G_r$ under which the set 
\begin{displaymath}   
I_{q} = \{ T^{A}_{\ BC}\,, R^{A}_{\ BCD}\,, T^{A}_{\ BC;M_{1}}\,, 
 R^{A}_{\ BCD;M_{1}}\,,T^{A}_{\ BC;M_{1}M_{2}}\,, \,\ldots, \, 
 R^{A}_{\ BCD;M_{1} \ldots M_{q}\,,} T^{A}_{\ BC;M_{1} \ldots M_{q+1}} \}  
\end{displaymath}     
is invariant, and the number $t_q$ of functionally independent 
functions of the space-time coordinates contained in $I_q$. 
They must be the same at each step $q\,$ ($0\leq q\leq p+1$) 
of the differentiation process
if the Riemann-Cartan manifolds are locally equivalent.

A practical procedure for testing equivalence of Riemann-Cartan 
space-times, which results from the above considerations,
starts by setting $q=0$ and has the following steps%
~\cite{frm,frm1,afmr1}: 
\begin{enumerate}
\item 
Calculate the set $I_{q}$
[the derivatives of the curvature up to the $q^{th}$ order 
and of the torsion up to the $(q+1)^{th}$ order].
\item 
Fix the frame, as much as possible, by putting
the elements of $I_{q}$ into canonical forms, and
find the residual isotropy group $H_{q}$ of transformations 
which leave these canonical forms invariant.
\item 
Find the number $t_{q}$ of functionally 
independent functions of space-time coordinates 
in the elements of $I_q$, brought into the canonical 
forms.
\item   
If the isotropy groups $H_{q}$ and $H_{(q-1)}$ 
are the same, and the number of functionally independent
functions $t_{q}$ is equal to $t_{(q-1)}$,
then let $q=p+1$ and stop. Otherwise, increment 
$q$ by 1 and go to step $1$.    
\end{enumerate}

To compare two Riemann-Cartan space-times we first test if
they have the same $t_q$ and $H_q$ for each $q$ up to $p+1$
[ $(p+2)^{th}$ derivative of the torsion ]. If they differ, so do
(locally) the Riemann-Cartan manifolds. If not, it is necessary
to check the consistency of eqs.\ (\ref{eqvcond}).

Since there are $t_p$  essential space-time coordinates
when the above procedure for testing equivalence terminates,
clearly $4-t_p$ are ignorable, so the isotropy group
will have dimension  $s = \mbox{dim}\,( H_p )$, and the group of 
symmetries (called affine isometries) of both metric (isometry) 
and torsion (affine collineations) will have dimension $r$ 
given by (see, for example, refs.~\cite{cartan}~--~\cite{frm1})
\begin{equation}
r = s + 4 - t_p \,, \label{gdim}
\end{equation}
acting on an orbit with dimension
\begin{equation}
d = r - s = 4 - t_p \,.  \label{ddim}
\end{equation}

In our implementation of the above practical procedure, 
rather than using the curvature and torsion tensors as such, the 
algorithms and computer algebra programs were devised and 
written in terms of spinor equivalents, namely~\cite{frm,frm1}:
(i) the irreducible parts of the Riemann-Cartan curvature, i.e.,
$\Psi_{ABCD}$,  $\Phi_{ABX'Z'}$, $\Theta_{ABX'Z'}$, $\Sigma_{AB}$,
$\Lambda$ and $\Omega$; and (ii) the irreducible parts of torsion, 
i.e., ${\cal T}_{AX'}$, ${\cal P}_{AX'}$ and ${\cal L}_{ABCX'}$. 

A relevant point to be taken into account when one
needs to compute derivatives of the curvature and the
torsion tensors is that they are interrelated by
the Bianchi and Ricci identities and their concomitants.
Thus, to cut down the number of quantities to be calculated
it is very important to find a set of quantities from
which the curvature and torsion tensors, and their covariant
derivatives are obtainable by algebraic operations. For 
Riemann-Cartan space-time manifolds, instead of using $I_{p+1}$ 
as such, we deal with a corresponding complete minimal set of  
quantities which are recursively defined in terms of totally 
symmetrized $q^{th}$ and $(q+1)^{th}\,$ (for $0\,\leq q\leq p+1\,$) 
derivatives of the curvature and torsion spinors, respectively%
~\cite{frm,frm1,fmr1}. In this work, however, we
shall only need the subsets of quantities for $q=0$ and $q=1$, which
can be taken to be (see~\cite{frm,frm1}) the quantities tabulated,
respectively, in the tables~\ref{tbi0} and~\ref{tbi1} below, where, 
to give an idea of the amount of calculations involved in the 
equivalence procedure, we have also included the number of real 
independent components of each spinorial quantities in the general 
(worst) case.
\begin{table}[htp] {\small 
\begin{center} 
\begin{tabular}{|l|c|c|}   
\hline  
{\sc tclassi}'s name & Spinor & Ind.\ Comp.  \\
\hline  \hline    
{\sc tpsi}     & $\Psi_{ABCD}$      &   10   \\   \hline
{\sc psiltor}  & $\psi_{ABCD}$      &   10   \\   \hline
{\sc tphi}     & $\Phi_{ABX'Y'}$    &    9   \\   \hline
{\sc philtor}  & $\phi_{ABX'Y'}$    &    9   \\   \hline
{\sc theta}    & $\Theta_{ABX'Y'}$  &    9   \\    \hline
{\sc dspttor}  & $\nabla^{(B}_{\ \ (B'}{\cal T}^{\ \ \ A)}_{X')}$ &  9 \\
\hline
{\sc dspptor}  & $\nabla^{(B}_{\ \ (B'}{\cal P}^{\ \ \ A)}_{X')}$ &  9 \\
\hline
{\sc sigma}    & $\Sigma_{AB}$      &      6  \\   \hline
{\sc bvttor}   & ${\cal M}_{AB}$    &      6  \\   \hline
{\sc bvptor}   & ${\cal B}_{AB}$    &      6  \\   \hline
{\sc spttor}   & ${\cal T}_{AX'}$   &    4   \\  \hline
{\sc spptor}   & ${\cal P}_{AX'}$   &    4   \\  \hline
{\sc tlambd}   & $\Lambda$          &      1  \\  \hline
{\sc omega}    & $\Omega$           &      1  \\  \hline
{\sc scttor}   & ${\cal T}$         &      1   \\  \hline
{\sc splttor}  & ${\cal L}_{ABCX'}$ &   16   \\  \hline
{\sc dspltor} & $\nabla^{(B}_{\ \ (B'}{\cal L}^{\ \ \ \ CDE)}_{X')}$ & 30  \\
\hline
\end{tabular}   
\end{center}      }
\caption{The 17 quantities from a complete minimal set for the step
$q=0$ of the equivalence algorithm
($I_0 = \{ T^A_{\ BC},\: R^A_{\ BCD},\: T^A_{\ BC;M_1} \} $ ).
{\sc tclassi}'s names and the number of real independent
components are also shown. In the general case there are a 
total of 140 real components.}
\label{tbi0}
\end{table}
\begin{table}[htp] {\small
\begin{center}
\begin{tabular}{|l|c|c|}
\hline  
{\sc tclassi}'s name & Spinor & Ind.\ Comp.  \\     
\hline  \hline 
{\sc dtpsi}   & $\nabla_{X'(E}\,\Psi_{ABCD)}$  & 24  \\  \hline 
{\sc dpsiltor}& $\nabla_{X'(E}\,\psi_{ABCD)}$  & 24  \\  \hline 
{\sc dtphi}   & $\nabla^{(C}_{\ \ (X'}\Phi^{AB)}_{\ \ \ \ Y'Z')}$ &  16 \\ 
\hline 
{\sc dphiltor}& $\nabla^{(C}_{\ \ (X'}\phi^{AB)}_{\ \ \ \ Y'Z')}$ & 16 \\
\hline 
{\sc dtheta}  & $\nabla^{(C}_{\ \ (X'}\Theta^{AB)}_{\ \ \ \ Y'Z')}$ &  16 \\
\hline 
{\sc d2spttor}  & 
$\nabla^{(C}_{\ \ (X'}\nabla^{A}_{\ \ Y'}{\cal T}^{B)}_{\ \ \ Z')}$
              &  16  \\     \hline 
{\sc d2spptor}  &
$\nabla^{(C}_{\ \ (X'}\nabla^{A}_{\ \ Y'}{\cal P}^{B)}_{\ \ \ Z')}$
              &  16  \\     \hline
{\sc dsigma}    & $\nabla_{X'(A}\,\Sigma_{BC)}$   & 16   \\  \hline 
{\sc dbvttor}   & $\nabla_{X'(A}\,{\cal M}_{BC)}$ & 16   \\  \hline 
{\sc dbvptor}   & $\nabla_{X'(A}\,{\cal B}_{BC)}$ & 16   \\  \hline 
{\sc aspttor}   & $\Box\, {\cal T}_{AX'}$        & 4   \\   \hline 
{\sc aspptor}   & $\Box\, {\cal P}_{AX'}$        & 4   \\   \hline 
{\sc dtlambd}   & $\nabla_{X'A}\,\Lambda$         &  4   \\  \hline 
{\sc domega}    & $\nabla_{X'A}\,\Omega$          &  4   \\  \hline 
{\sc dscttor}   & $\nabla_{X'A}\,{\cal T}$        &  4   \\  \hline 
{\sc d2spltor}  & 
$\nabla^{(A}_{\ \ (X'}\nabla^{B}_{\ Y'}{\cal L}^{\ \ \ CDE)}_{Z')}$
             & 48     \\     \hline
{\sc aspltor}   & $\Box\, {\cal L}_{X'ABC}$      & 16  \\   \hline
{\sc txi}       & $\Xi_{X'ABC}$                & 16  \\  \hline
{\sc xith}      & ${\cal X}_{X'ABC}$           & 16  \\  \hline
{\sc tsigm}     & ${\cal U}_{AX'}$             & 4   \\  \hline
{\sc psigm}     & ${\cal V}_{AX'}$             & 4   \\   
\hline
\end{tabular}
\end{center}        }  
\caption{The 21 quantities which, together with the 17 quantities 
of table 1, form a complete minimal set for the step
$q=1$ of the equivalence algorithm
($I_{1}= I_{0} \cup \{R^{A}_{\ BCD;M_{1}}, T^{A}_{\ BC;M_{1}M_{2}}\}$).
{\sc tclassi}'s names and the number of real independent
components are also shown. These 21 quantities have a total of 300 
real components in the general case.}
\label{tbi1}
\end{table}
\newpage

To close this section we remark that, in line with the
usage in the literature, in the {\sc tclassi} implementation 
of the above results it is used a notation in 
which the indices are all subscripts, components are labelled
by a primed and unprimed index whose numerical values are the
sum of corresponding (prime and unprimed) spinors indices. 
Thus, for example, one has 
$\nabla\,\Psi_{20'} \equiv  \Psi_{(1000;1)0'}$\,,
where the parentheses indicate symmetrization.

\vspace{6mm}
{\raggedright
\section{Main Results and Conclusions}  }
\setcounter{equation}{0} 

The basic idea behind our procedure for checking the local 
equivalence of RC space-times, discussed in the previous section, 
is a separate handling of frame rotations and space-time coordinates, 
fixing the frame at each stage of differentiation (of the curvature 
and torsion tensors) by aligning the basis vectors as far as possible 
with invariantly-defined directions. This is done in practice, 
by bringing to canonical forms first the quantities with the 
same symmetry as the Weyl spinor (the Weyl-type spinors: $\Psi_A$ 
and $\psi_A$) followed by the spinors with the symmetry of the 
Ricci spinor (the Ricci-type spinors: $\Phi_{AB'}$, $\phi_{AB'}$, 
$\Theta_{AB'}$, $\nabla {\cal T}_{AX'}$, $\nabla {\cal P}_{AY'}$), then
bivector spinors ($\Sigma_{AB}$, ${\cal M}_{AB}$, ${\cal B}_{AB}$), 
and finally vectors (${\cal T}_{AX'}$, ${\cal P}_{AX'}$) are taken 
into account. 
Thus, if $\Psi_{A}$ is Petrov type D, for example, the frame is fixed
by demanding that the only nonvanishing component of $\Psi_A$
is $\Psi_2$. On the other hand, if $\Psi_{A}$ is Petrov type I the 
frame can be fixed by requiring that the components of 
$\Psi_A$ are such that $\Psi_1 = \Psi_3 \not= 0, \Psi_2 \not= 0$. 
Clearly an alternative canonical frame for Petrov type I 
is obtained by imposing $\Psi_0 = \Psi_4 \not= 0, \Psi_2 \not= 0$. 
Although the latter choice is implemented in {\sc tclassi} 
as the canonical frame for Petrov type I, in this section we 
shall use the former (defined to be an acceptable alternative 
in {\sc tclassi}) to make straightforward the comparison 
of our findings with the previous results on G\"odel-type 
space-times~\cite{rebaman,AmanFonsecaMacCallumReboucas97}.  

We shall consider now a class of four-dimensional Riemann-Cartan 
manifolds $M$, endowed with a G\"odel-type metric (\ref{ds2}) and 
a torsion that is aligned with the preferred direction defined by 
the rotation vector field, but which does not share the same 
translational symmetries of the metric (\ref{ds2}). Actually the 
class of RC space-times we are concerned with here is such that in 
the coordinate system in which (\ref{ds2}) is given, the 
nonvanishing components of the torsion reduce to 
\begin{equation}   \label{torcomp}
T^t_{\ xy} \equiv D(x)\, S(z)\,.
\end{equation}
It should be emphasized that in the Lorentz frame relative to
which the G\"odel-type line element~(\ref{ds2}) reduces to
\begin{equation} \label{ds2f}
ds^2 =\eta_{AB}\,\,\omega^{A}\, \omega^B  \qquad \mbox{with} \qquad
\eta_{AB}={\rm diag}\,(+1,-1,-1,-1)\,, \: 
\end{equation}
and 
\begin{equation} \label{lort}
\omega^{0} = dt + H(x)\,dy\,, \qquad
\omega^{1} = dx\,, \qquad
\omega^{2} = D(x)\,dy\,, \qquad
\omega^{3} = dz\,,          
\end{equation}
the only novanishing component of the torsion is
\begin{equation}   \label{torcompf}
T^0_{\ 12} = S(z)\,.
\end{equation}
Therefore, the function $D(x)$ in (\ref{torcomp}) and 
in expression for the torsion in {\AA}man {\em et al.\/}%
~\cite{AmanFonsecaMacCallumReboucas97} can be eliminated by a 
suitable choice of basis.
 
For arbitrary functions $H(x)$, $D(x)$ and $S(z)$, the Weyl-type
spinor $\Psi_A$ is Petrov type I, whereas $\psi_A$ is Petrov type D; 
this fact can be easily checked by using the module {\sc segpet}
of {\sc tclassi}. 
Accordingly the null tetrad $\theta^A$ which turns out to be 
appropriate (canonical) for our purpose here is

\parbox{14cm}{\begin{eqnarray*} 
\theta^{0} =\frac{1}{\sqrt{2}}\left[dt + H(x)\,dy + dz\,\right]\,,\qquad
\theta^{2} =\frac{1}{\sqrt{2}}\left[D(x)\,dy - i\,dx\,\right]\,, \\
\theta^{1} =\frac{1}{\sqrt{2}}\left[dt + H(x)\,dy - dz\,\right]\,, \qquad
\theta^{3} = \frac{1}{\sqrt{2}}\left[D(x)\,dy + i\,dx\,\right]\,. 
            \end{eqnarray*}}  \hfill
\parbox{1cm}{\begin{eqnarray} \label{nullt}  \end{eqnarray}}
 
Clearly in this basis the G\"odel-type line element (\ref{ds2}) and 
the torsion tensor $T$ are, respectively, given by 
\begin{equation}
ds^2 = 2\,(\theta^0\,\theta^1 - \theta^2\,\theta^3) \qquad   \label{gtyrc}
\mbox{and} \qquad                      
T^0_{\ 23} = T^{1}_{\ 23} = \frac{\sqrt{2}}{2}\,i\,\,S(z)\,.
\end{equation}

\begin{sloppypar}
It is worth mentioning that the Petrov type for $\Psi_A$ and $\psi_A$
and the canonical frame~(\ref{nullt}) were obtained by interaction with
{\sc tclassi}, starting from a Lorentz frame, changing
to a null tetrad frame, and making dyad transformations to bring
$\Psi_A$ and $\psi_A$, respectively, into the canonical form for 
Petrov types I and D. As a matter of fact, in the frame~(\ref{nullt}) 
all quantities of the  sets $I_0$ and $I_1$ are brought into their 
corresponding canonical forms.
\end{sloppypar}

Using the {\sc tclassi} package one finds the following nonvanishing 
components of the quantities in the complete minimal set corresponding 
to $I_0$ (see table~\ref{tbi0}) of our algorithm:

\begin{eqnarray}
\Psi_1 &=& \Psi_3 = - \frac{1}{8}\, \left( \frac{H'}{D}\, \right)' \,,
                           \label{1st}  \\
\Psi_2 &=& - \,\frac{S}{4}\, \left(\frac{S}{3} - \frac{H'}{D}\, \right) 
+\frac{1}{6}\left[\,\frac{D''}{D} - \left( \frac{H'}{D}\, \right)^2\, 
           \,\right] + \, \frac{i}{6}\, \dot{S} \,, \\  
\psi_2 &=& - \,\frac{S}{4}\,\left( S - \frac{H'}{D}\, \right) +
                     \, \frac{i}{6}\, \dot{S} \,, \\  
\Phi_{00'}&=&\Phi_{22'} = \frac{S}{4}\, \left(\frac{S}{2} 
-\,\frac{H'}{D}\, \right) +\frac{1}{8}\, \left(\frac{H'}{D}\,\right)^2 \,,\\ 
\Phi_{01'}&=&\Phi_{12'} = \frac{1}{8}\, \left(\frac{H'}{D}\,\right)' \,, \\
\Phi_{11'}&=& \,\frac{S}{4}\,\left(\,\frac{S}{4} - \frac{H'}{D}\, \right)  
  + \frac{1}{4} \left[\, \frac{3}{4}\,\left(\frac{H'}{D}\,\right)^2 
  - \,\frac{D''}{D} \right] \,, \\
\phi_{00'}&=& \phi_{22'} =\: \phi_{11'} = \frac{S}{4}\, 
                     \left( S - \frac{H'}{D}\,\right) \,, \\ 
\Theta_{00'} &= &\Theta_{22'} =\: 2\,\Theta_{11'}= \frac{\dot{S}}{4}\,, \\
\nabla\,{\cal P}_{00'} &=& \nabla\,{\cal P}_{22'}=
          \:-2\,\nabla\,{\cal P}_{11'}= -\,\frac{\dot{S}}{2} \,, \\
{\cal P}_{00'}&=& - {\cal P}_{11'} = - \,\frac{\sqrt{2}}{2}\,\, S \,, \\ 
\Lambda &=& - \, \frac{S^2}{48}\,- \frac{1}{12} \left[\,\frac{D''}{D}
   - \frac{1}{4} \left(\frac{H'}{D}\,\right)^2  \right] \,,  \\
\Omega &=& \,\frac{\dot{S}}{24} \,, \\
{\cal L}_{10'}&=& {\cal L}_{21'} = - \,\frac{i}{6}\,\sqrt{2}\,\,S \,, \\
\nabla {\cal L}_{10'}&=&- \,\nabla {\cal L}_{32'}= \frac{S}{16}\,
                \left( S -\frac{H'}{D}\, \right)
               - \, \frac{i}{8}\, \dot{S} \,, \label{last}
\end{eqnarray}
where the prime and overdot denote, respectively, derivative with 
respect to $x$ and $z$.
Note, incidentally, that $\psi_A$ is Petrov type D for this class 
while for the RC G\"odel-type space-times discussed in~\cite{afmr} 
both Weyl-type spinor $\Psi_A$ and $\psi_A$ are
Petrov type I, making clear that the underlying
RC manifolds are not (locally) equivalent, as one may have
noticed from the outset.

Before proceeding to the second step of our practical procedure
let us introduce a notation. In line with the usage in the literature
(see, for example, \cite{PaivaReboucasMacCallum1993}) we shall refer 
to the spinorial (frame) components of the quantities in the sets $I_q$ 
(and the quantities of the corresponding complete minimal sets)
for each step $q$ of the equivalence problem algorithm as {\em Cartan 
scalars\/}, since they are scalars under coordinate transformations.

Following the algorithm of the previous section, one needs to find
the isotropy group which leaves the above Cartan scalars (canonical
forms) invariant, and the number of functionally independent functions
of the space-time coordinates among these Cartan scalars. One can easily
find that the above whole set of Cartan scalars%
~(\ref{1st})~--~(\ref{last}) is not invariant under any subgroup
of the Lorentz group, so $\mbox{dim}\,( H_0 ) = 0$. Moreover, for
arbitrary smooth real functions $H(x)$, $D(x)$ and $S(z)$ the
number of functionally independent functions of the coordinates 
in the complete minimal set corresponding to $I_0$ clearly is $t_0=2$.

The next steps of our algorithm are (i) to calculate the Cartan 
scalars of table~\ref{tbi1}, (ii) to find the residual isotropy 
group which leaves these quantities invariant, and (iii) to find 
if there is any additional functionally independent function of 
the coordinates in the set of Cartan scalars of table~\ref{tbi1}. 
For the sake of brevity we shall present the results without going 
into details of the calculations, which can be easily worked out
by using {\sc tclassi}. The nonvanishing Cartan 
scalars of table~\ref{tbi1} are {\sc dtpsi}, {\sc dpsiltor},
{\sc dtphi}, {\sc dphiltor}, {\sc dtheta}, {\sc d2spltor}, 
{\sc d2spptor}, {\sc aspltor}, {\sc aspptor}, {\sc dtlambd},
{\sc domega}, {\sc txi}, and {\sc xith}. These scalars
are invariant under no subgroup of the Lorentz group 
[$\,\mbox{dim}\,(H_0) = \mbox{dim}\, (H_1)= 0\,$], and are such 
that $t_1= t_0 = 2$. Thus, no new covariant derivative 
should be calculated.
{}From eq.~(\ref{gdim}) one finds that the group of symmetries 
(affine-isometric motions) of this class of Riemann-Cartan 
G\"odel-type space-times is two-dimensional.

We shall now consider the class of Riemann-Cartan G\"odel-type 
space-time manifolds where the underlying {\em Riemannian\/} manifolds 
are ST homogeneous~\cite{rebtio,rebaman}, i.e., when the conditions
(\ref{metcond1}) and (\ref{metcond2}) hold. In this case
the nonvanishing Cartan scalars corresponding to the first 
step of our algorithm for $q=0$ reduce to
\begin{eqnarray}  
\Psi_2 &=& \, \frac{S}{2}\, \left(\,\omega - \frac{S}{6} \,\right)
           + \frac{m^2}{6} - \frac{2}{3}\,\omega^2 +
            \, \frac{i}{6}\,\dot{S} \,, \label{um} \\
\psi_2 &=& - \, \frac{S}{4}\,\left(S -2\,\omega\,\right)
            + \,\frac{i}{6}\,\dot{S}  \,,\label{dois} \\ 
\Phi_{00'}&=&\Phi_{22'} =\frac{S}{4}\,\left(\frac{S}{2} 
 -2\,\omega\,\right) +\frac{\omega^2}{2} \,, \label{tres} \\
\Phi_{11'}&=& \,\frac{S}{4}\,\left(\,\frac{S}{4} - 2\,\omega\, \right)  
  +  \frac{3}{4}\,\omega^2 - \frac{m^2}{4} \,, \label{quatro} \\
\phi_{00'}&=& \phi_{22'} = \: \phi_{11'} = \,\frac{S}{4}\, 
                \left( S - 2\,\omega\,\right) \,, \label{cinco} \\
\Theta_{00'} &= &\Theta_{22'} =\: 2\,\Theta_{11'}= 
                \, \frac{\dot{S}}{4}\,, \label{seis} \\
\nabla {\cal P}_{00'} &=& \nabla {\cal P}_{22'}=
    - \:2\,\nabla {\cal P}_{11'}= - \,\frac{\dot{S}}{2} \,, \label{sete} \\
{\cal P}_{00'}&= & -{\cal P}_{11'} =
            -\,\frac{\sqrt{2}}{2}\,\, S\,,\label{oito} \\ 
\Lambda &=&-\,\,\frac{S^2}{48}\,+\frac{1}{12} \left(\,\omega^2-m^2 \right)\,,
                                        \label{nove} \\
\Omega &=& \,\frac{\dot{S}}{24} \,, \label{dez} \\
{\cal L}_{10'}&=& {\cal L}_{21'} =
              -\,\frac{i}{6}\,\,\sqrt{2}\,S\,, \label{onze} \\
\nabla {\cal L}_{10'}&=&  - \,\nabla {\cal L}_{32'}= \,\frac{S}{16}\,
                \left( S - 2\,\omega\,\right)
          - \,\frac{i}{8}\,\dot{S} \,. \label{doze}
\end{eqnarray}

Following the algorithm of the previous section, one needs to find
the isotropy group which leaves the above Cartan scalars (canonical
forms) invariant as well as the number of functionally independent
functions of the coordinates among these Cartan scalars. 
Clearly as $S$ is an arbitrary smooth real function one has $t_0=1$. 
As far as the isotropy group $H_0$ is concerned, since $S\not= 0$ one can 
easily find that although there are Cartan scalars as, e.g., $\Omega$ 
and $\Lambda$ which are invariant under the Lorentz group, the 
whole set of Cartan scalars (\ref{um})~--~(\ref{doze}) is invariant 
only under the subgroup of spatial rotation
\begin{equation} \label{SpaRot}
 \left[
\begin{array}{cc}
   e^{i\alpha} &      0       \\
        0      & e^{- i\alpha} \\
\end{array} 
\right] \,\,,
\end{equation}
where $\alpha$ is a real parameter. So, the isotropy group $H_0$
is such that  $\mbox{dim}\: H_0= 1$. 

For the following step ($q=1$) of our algorithm one readily finds
\begin{eqnarray} 
\nabla \,\Psi_{20'} &=& -\nabla \,\Psi_{31'} =
\frac{i}{40}\,\sqrt{2}\,\,S \left(2\,m^2 -20\,\omega^2  
           +8\,\omega S\, - S^2  \,\right)   \nonumber \\
&&
+\,\frac{i}{10}\,\sqrt{2}\,\omega \left( 4\,\omega^2 - m^2\right)
+\,\frac{\sqrt{2}}{20}\,\,\left[\, (5\,\omega - 2\,S )\,\dot{S}
                    + i\,\ddot{S}\,\right] \,, \label{primeiro} \\
\nabla \,\psi_{20'} &=& -\nabla \,\psi_{31'} =
 -\frac{i}{40}\,3\,\sqrt{2}\,\,S \left( 4\,\omega^2
   - 4\,\omega\,S + S^2 \,\right) \nonumber \\
&&   
+\,\frac{\sqrt{2}}{20}\,\left[\,(5\,\omega-4\,S) \dot{S} 
             + i\,\ddot{S} \,\right] \,,  \\
\nabla \,\Phi_{00'} &=& -\,\nabla\,\Phi_{33'} =
\frac{\sqrt{2}}{8}\, \left(S - 2\,\omega \right) \,\dot{S} \,,  \\
\nabla \,\Phi_{11'} &=& - \,\nabla\,\Phi_{22'} =
\frac{\sqrt{2}}{72}\, \left(S - 6\,\omega \right) \,\dot{S} \,,  \\
\nabla \,\phi_{00'} &=& -\,\nabla\,\phi_{33'}\, =
\,3\,\nabla \,\phi_{11'}  = - \,3\,\nabla\,\phi_{22'} =
\frac{\sqrt{2}}{4}\, \left(S - \omega \right) \,\dot{S} \,, \\
\nabla \,\Theta_{00'} &=& -\,\nabla\,\Theta_{33'}\:\, =
\:9\,\nabla \,\Theta_{11'}\:\, = \:-\,9\,\nabla\,\Theta_{22'}\:\,=
\frac{\sqrt{2}}{8}\,\, \ddot{S} \,,  \\         
\nabla^2\, {\cal P}_{00'}&=& - \,\nabla^2\, {\cal P}_{33'} = 
-\,3\,\nabla^2\, {\cal P}_{11'} =  \,3\,\nabla^2\, {\cal P}_{22'} = 
            \, - \frac{\sqrt{2}}{4}\,\,\ddot{S} \,,   \\ 
\Box\, {\cal P}_{00'} &=& -\,\Box\, {\cal P}_{11'} = 
                  \frac{\sqrt{2}}{2}\,\, \ddot{S} \,, \\
\nabla \, \Lambda_{00'} &=&-\, \nabla \, \Lambda_{11'} = -\, 
                   \frac{\sqrt{2}}{48}\,S\, \dot{S} \,, \\
\nabla \, \Omega_{00'} &=& -\,\nabla \, \Omega_{11'} = \, 
                  \frac{\sqrt{2}}{48}\,\, \ddot{S} \,, \\
\nabla^2 {\cal L}_{10'}&=& \,\nabla^2 {\cal L}_{43'}= 
 \frac{i}{80}\,\sqrt{2}\,\,S \left( 4\,\omega^2 
   - 4\,S\, \omega + S^2 \,\right) \nonumber \\
&& 
  + \frac{1}{40}\,\sqrt{2}\,\, \left[\, (3\,S\,- 4\,\omega)\,\dot{S} 
                  - 2\,i\,\ddot{S} \,\right] \,, \\ 
\nabla^2 {\cal L}_{21'}&=& \,\nabla^2 {\cal L}_{32'}= 
-\,\frac{i}{480}\,\sqrt{2}\,\,S \left(4\,\omega^2 
   - 4\,S\, \omega +S^2 \,\right) \nonumber \\
&&   
  + \frac{1}{240}\,\sqrt{2}\,\, \left[\, (4\,\omega - 3\,S)\,\dot{S} 
              + 2\,i\,\ddot{S} \,\right] \,, \\ 
\Box\,{\cal L}_{10'}&=& \,\Box\,{\cal L}_{21'}= 
\frac{i}{4}\,\sqrt{2}\,\,S \left(4\,\omega^2 
               - 4\,\omega \,S + S^2 \,\right) 
             + \frac{i}{6}\,\sqrt{2}\,\,\ddot{S} \,,   \\ 
\Xi_{10'} & = & \Xi_{21'} =
 \frac{i}{16}\,\sqrt{2}\,\,S \left( 2\,m^2 + 8\,S\, \omega 
   - S^2 - 20\, \omega^2\, \right) \nonumber \\
&&
+\frac{i}{4}\,\sqrt{2}\,\omega\left( 4\,\omega^2 - m^2 \right) 
-\,\frac{\sqrt{2}}{24} \left(S\,\dot{S}+2\,i\,\ddot{S}\right) \,, \\
{\cal X}_{10'} & = & {\cal X}_{21'} =                            
-\,\frac{\sqrt{2}}{24} \left[\,(3\,S - 6\,\omega)\,i\,\dot{S}
+2\,\ddot{S}\right] \label{ultimo} \,.
\end{eqnarray} 

As no new functionally independent function came out, then 
$t_0=t_1 =1$. Besides, the Cartan scalars 
(\ref{primeiro})~--~({\ref{ultimo}) are invariant under 
the same isotropy group (\ref{SpaRot}), i.e. $H_0 = H_1$. 
Thus no new covariant derivative should be calculated. 
{}From eq.~(\ref{gdim}) one finds that the group of symmetries 
(affine-isometric motions) of this particular class of 
Riemann-Cartan G\"odel-type space-times is four-dimensional. 
Note, however, that these RC manifolds are not ST homogeneous
(see below for a formal definition of ST homogeneity) 
as one might think at a first sight. Indeed, from eq.~(\ref{ddim})
one readily find that the orbit of an arbitrary point 
$P$ on a manifold of this class, under the action of the group 
of symmetries, is three-dimensional.

We shall finally focus our attention in the ST homogeneous
RC G\"odel-type space-times. A word of clarification in order here: 
a $n$-dimensional Riemann-Cartan manifold $M$ is said to
be homogeneous when the orbit of an arbitrary point $P \in  M$
under the action of the group of affine-isometric motions $G_r$ is
the manifold $M$ itself. Clearly for ST homogeneity of a
four-dimensional RC manifold we have an orbit of dimension $d=4$. Now, from 
equation~(\ref{ddim}) one  finds that for ST homogeneity ($d=4$) we 
must have $t_{q} = 0$, for all steps $q$ in the equivalence procedure.
Thus, from  eqs.\ (\ref{1st}) -- ({\ref{last}) one easily 
concludes that for the class of Riemann-Cartan G\"odel-type 
space-times (\ref{gtyrc}) [ or equivalently given by a 
metric (\ref{ds2f}) and (\ref{lort}), and a torsion 
(\ref{torcompf}) ] to be ST homogeneous, besides
the conditions (\ref{metcond1})~--~(\ref{metcond2})
it is necessary that it has a constant torsion, i.e., 
that the condition 
\begin{equation}
S=const \equiv S_0      \label{torcond}
\end{equation}
holds. These necessary conditions are also sufficient for ST 
homogeneity of our class of RC G\"odel-type manifolds. 
Indeed, under the conditions (\ref{metcond1}), (\ref{metcond2})
and (\ref{torcond}) the Cartan scalars 
(\ref{um})~--~(\ref{doze}) and (\ref{primeiro})~--~(\ref{ultimo}) 
corresponding to the first and second steps of differentiation
in our algorithm
reduce, respectively, to eqs.\ (3.23)~--~(3.31) and 
eqs.\ (3.33)~--~(3.38) of~\cite{AmanFonsecaMacCallumReboucas97}, 
which imply that the corresponding  RC G\"odel-type space-times 
are ST homogeneous with a $G_5$ of symmetry, and
characterized by three independent parameters $S_0$, $m^2$ and $\omega$:
[identical triads $(S_0, m^2, \omega)$ specify locally equivalent RC 
G\"odel-type space-time manifolds].
As a matter of fact, from eqs.\ (\ref{um})~--~(\ref{doze}) and 
(\ref{primeiro})~--~(\ref{ultimo}), by using the equivalence problem 
techniques, it is straightforward to show that (\ref{metcond1}),
(\ref{metcond2}) and (\ref{torcond}) are the necessary and 
sufficient conditions for ST homogeneity of these RC G\"odel-type
manifolds, which admit a $G_5$ of symmetries.

The above results for the ST homogeneous G\"odel-type
class of RC space-time manifolds extend the theorems
1 and 2 of Ref.~\cite{AmanFonsecaMacCallumReboucas97} 
(given below, for completeness) to the case in which the torsion 
although aligned along the direction of the rotation is 
does not share the same translational 
symmetry of the G\"odel-type metric~(\ref{ds2}). The above-mentioned 
generalizations of the theorems 1 and 2 can be stated as follows:

\vspace{2mm} 
\begin{theorem} \label{HomCond}
The necessary and sufficient conditions for a Riemann-Cartan
space-time manifold $M$ endowed with a G\"odel-type  
metric (\ref{ds2f}) and (\ref{lort}), and a torsion 
(\ref{torcompf}) [ or equivalently with metric (\ref{ds2})
and torsion (\ref{torcomp}) ]
to be ST locally homogeneous~\footnote{
For a clear and formal distinction between
local and global (topological) homogeneity of a manifold see,
for example, Koinke {\em et al.\/}~\cite{KoikeTamimotoHosoya94}.} 
are given by  equations 
(\ref{metcond1}), (\ref{metcond2}) and (\ref{torcond}).
\end{theorem}
\begin{theorem} \label{RelPar}   \begin{sloppypar}
All ST locally
 homogeneous Riemann-Cartan 
space-time manifolds $M$ endowed with a G\"odel-type  
metric (\ref{ds2f}) and (\ref{lort}), and a torsion 
(\ref{torcompf}) [ or equivalently with metric (\ref{ds2}) 
and torsion (\ref{torcomp}) ]
admit a five-dimensional group of affine-isometric motions and
are characterized by three independent parameters 
$m^2$, $\omega$ and $S_0$: identical triads ($m^2, \omega, S_0$) 
specify locally equivalent manifolds.     \end{sloppypar}
\end{theorem}

It should also be stressed that when $S = 0$ and the conditions 
(\ref{metcond1})~--~(\ref{metcond2}) hold the Cartan scalars 
(\ref{um})~--~(\ref{doze}) and (\ref{primeiro})~--~(\ref{ultimo}) 
reduce to the corresponding scalars for {\em Riemannian\/} 
G\"odel-type space-times 
(eqs.\ (3.12)~--~(3.15) and (3.18)~--~(3.21) in~\cite{rebaman}). 
Therefore, the results in~\cite{rebaman} can be reobtained 
as a particular case of our study in this work.

It it is worth emphasizing that by the procedure 
to test local equivalence we have used throughout this work,
we actually compute one invariant local characterization 
of each class of Riemann-Cartan space-times, and at the 
end of the procedure in addition to $t_q$'s and 
$H_q$'s we have a number of consequent data as, for example,
the dimension of the symmetry group [given by (\ref{gdim})],
the dimension of the orbit of an arbitrary point under its 
action [given by (\ref{ddim})], and the algebraic classifications 
(Petrov and Segre) of Weyl-type and Ricci-type spinors 
(needed to fix the frame at the step $q=0$ of our algorithm). 
Furthermore, the complete set of Cartan scalars $I_{p+1}$ 
for each class or RC G\"odel-type space-time manifolds 
give a (local) coordinate-invariant description of the 
gravitational field in each class of RC G\"odel-type
manifolds irrespective of the torsion theory of 
gravitation one may be concerned with (for a list
of G\"odel-type solutions to the Einstein-Cartan field
equations, for example, see~\cite{SinghAgrawal94}). This reveals
the importance of our results in the general context 
of the torsion theories of gravitation in which
Riemann-Cartan manifolds are the underlying arena for
the formulation of the physical laws.

To close this article in what follows we summarize the main 
results we have obtained.
By using the equivalence techniques embodied in the suite of 
computer algebra programs (called {\sc tclassi})
we have examined a class of Riemann-Cartan space-time manifolds
$M$ endowed with a G\"odel-type  metric (\ref{ds2f}) with 
(\ref{lort}), and a torsion (\ref{torcompf}) [ or equivalently 
with metric (\ref{ds2}) and torsion (\ref{torcomp}) ].
We have shown that in the general case, i.e.,  when $H(x), D(x)$ and 
$S(z)$ are arbitrary smooth real functions these RC space-times admit 
a group $G_{2}$ of affine-isometric motions. On the other hand,
when the conditions for ST homogeneity (\ref{metcond1}) and 
(\ref{metcond2}) of the underlying Riemannian manifold are imposed 
the resulting family of Riemann-Cartan G\"odel-type permits a group
$G_4$ of affine-isometric motions. 
Moreover, when besides the conditions 
(\ref{metcond1})~--~(\ref{metcond2}) one has a constant torsion 
[ given by (\ref{torcompf}) with $S(z)=const$ ] the group of 
affine-isometric motions of these Riemann-Cartan G\"odel-type 
manifolds is five-dimensional.
We have also derived the above theorems 1 and 2 for
the case in which the torsion is polarized along 
the direction of the rotation but does not share the same 
translational symmetries of the metric (\ref{ds2}),
extending, therefore, the results found by {\AA}man
{\em et al.\/}~\cite{AmanFonsecaMacCallumReboucas97}.
Finally, the results of the Riemannian G\"odel-type
manifolds found by Rebou\c{c}as and {\AA}man~\cite{rebaman}
have been recovered in the limit when the torsion
vanishes.

\end{document}